\documentclass[%
reprint,
 amsmath,amssymb,
 aps,
]{revtex4-2}

\usepackage{graphicx}

\graphicspath{ {./figs/} }

\begin{document}

\title{Propagation of Picosecond Pulses on Superconducting Transmission Line Interconnects}

\author{Vladimir V. Talanov\textsuperscript{1}, Derek Knee\textsuperscript{2}, David Harms\textsuperscript{1}, Kieran Perkins\textsuperscript{1}, Andrew Urbanas\textsuperscript{1}, Jonathan Egan\textsuperscript{1}, Quentin Herr\textsuperscript{1}, and Anna Herr\textsuperscript{1}}
\affiliation{\textsuperscript{1}Northrop Grumman Corp., Baltimore, MD 21240}
\affiliation{\textsuperscript{2}Microsoft Research, Redmond, WA 98052}

\thanks{This research is based upon work supported in part by the
  ODNI, IARPA, via ARO. The views and conclusions contained herein are
  those of the authors and should not be interpreted as necessarily
  representing the official policies or endorsements, either expressed
  or implied, of the ODNI, IARPA, or the U.S. Government.}

\date{14 April 2021}

\begin{abstract}
Interconnects are a major discriminator for superconducting digital
technology, enabling energy-efficient data transfer and high-bandwidth
heterogeneous integration. We report a method to simulate propagation
of picosecond pulses in superconducting passive transmission lines
(PTLs). A frequency-domain propagator model obtained from the Ansys
High Frequency Structure Simulator (HFSS) field solver is incorporated
in a Cadence Spectre circuit model, so that the particular PTL
geometry can be simulated in the time-domain. The Mattis-Bardeen
complex conductivity of the superconductor is encoded in the HFSS
field solver as a complex-conductivity insulator. Experimental and
simulation results show that Nb 20\,$\Omega$ microstrip PTLs with
1\,$\mu$m width can support propagation of a single-flux-quantum
pulse up to 7\,mm and a double-flux-quantum pulse up to 28\,mm.
\end{abstract}

\maketitle

Low-loss interconnects are a major advantage for superconducting
digital technology as they support serial data links at the on-chip
clock rate and with the on-chip signal levels. This enables off-die
communication with low overhead
\cite{herr2002high}-\nocite{hashimoto2005demonstration}\cite{filippov2017experimental}.
The transmission energy in superconducting passive transmission lines
(PTLs) links is on order fJ/bit including the 300$\times$ cooling
overhead. This is four orders of magnitude lower than energy-efficient
CMOS interconnects reporting 0.5-2\,pJ/bit inter-chiplet communication
\cite{farjadrad2019bunch} and 0.1-1\,nJ/bit for server-to-server
communication \cite{miller2017attojoule}. Superconducting links share
the essential properties of optical interconnect but without the
overhead of low conversion efficiency.

Starting from the demonstration of 60\,Gb/s chip-to-chip
communication, superconducting off-die PTL links have advanced to
demonstration of 16-bit buses \cite{filippov2017experimental} using a
single chip on an multi-chip module (MCM), and the demonstration of an
MCM with multiple chips of different sizes and with synchronous
communication between them \cite{Egan2020}. This design used an
accurate model for the PTL interconnects as presented in this paper,
enabling simulation and verification methods compatible with industry
standard CAD tools.

Accurate PTL modeling must capture the attenuation and dispersion of
signals with analog bandwidth extending to sub-THz. The energy
efficient superconducting logic families
\cite{herr2011ultra},\cite{Volkmann2015} encode data in a form of
single-flux-quantum (SFQ) pulses of picosecond duration. Only quantum
flux parametrons (QFPs) are incompatible with PTL interconnect,
requiring SFQ signal translation for transmit and receive
\cite{china2016demonstration}.

The SFQ pulse spectrum, 360\,GHz for Josephson junctions with critical
current density of 100\,$\mu$A/$\mu$m$^2$, approaches the material
energy-gap frequency of 720\,GHz for Nb interconnect. Previous
simulation studies have modeled the data links using a two-fluid model
\cite{rafique2005optimization}, \cite{TAKEUCHI2009}. The model is
applicable up to 100\,GHz, for PTLs with dielectric that is thick
compared to the magnetic penetration depth, $\lambda$. Short-range
propagation of SFQ pulses on PTLs has been confirmed in numerous
experiments,
e.g. \cite{Semenov93}-\nocite{TanakaPTL2009}\nocite{LincolnMCM}\cite{shukla2019investigation}
but accurate modeling is required for future systems with sub-micron,
high-impedance PTLs covering distances up to 1\,meter.

Kautz was the first to accurately model the propagation of picosecond
pulses on superconducting transmission line with a numerical approach
incorporating the Mattis-Bardeen microscopic theory
\cite{mattis1958theory} for the superconductor complex conductivity
$\sigma=\sigma_1-i\sigma_2$ \cite{kautz1978picosecond}. Shortly after
Peterson and McDonald reported the generation of such pulses in the
Josephson junction \cite{peterson1977picosecond}, Kautz considered a
parallel-plate transmission line with Nb groundplane and Pb-In
stripline, and showed that dispersion both attenuates and distorts the
pulse while it propagates. Chi et al.\ verified Kautz's result
qualitatively by electro-optic sampling of picosecond pulses in a Nb
coplanar transmission line \cite{chi1987subpicosecond}, and comparing
to the model.

Kautz's approach applies to a wide microstrip PTL where the
fringe-fields can be ignored. Real PTLs, however, may include any
number of complications, as they may be microstrips whose width is
comparable to the dielectric thickness, may be coplanar or stripline,
may be in proximity to via walls or other transmission lines, may
utilize an insulator with non-uniform dielectric constant, and may
incorporate discontinuities such as bends, vias, or crossings
\cite{rafique2005optimization}. A 3D field solver such as Ansys HFSS
\cite{HFSS} allows accurate modeling of all of this, provided an
accurate frequency-dependent model of the dispersive material.

Here we report a method to simulate propagation of picosecond pulses
in superconducting PTLs with arbitrary length and geometry.  The
method uses a frequency domain propagator model obtained from the
Ansys HFSS field solver to produce a Cadence Spectre circuit model,
for time-domain simulation. In the frequency-domain model, the
scattering matrix $\hat{S}(\omega)$ between input and output ports of
the PTL is obtained using the Mattis-Bardeen complex
conductivity. $\hat{S}(\omega)$ is converted into the impulse response
matrix $\hat{S}(t)$ to produce the time-domain model, implemented in
Verilog-A. This model is used to simulate and optimize the PTL link
including the reciprocal quantum logic (RQL) driver and receiver. The
models are verified by comparing to the known theoretical solutions
for a wide microstrip \cite{kautz1978picosecond} and by comparing to
experimental data.

\section{General Method}

The frequency-domain and time-domain responses of any multiport
network can be related by \cite{shlepnev2010quality}
\begin{align*}
  & \hat{b}(\omega)=\hat{a}(\omega)\,\hat{S}(\omega),
  & \hat{b}(t)=\int_{-\infty}^\infty \hat{a}(\tau)\,\hat{S}(t-\tau)\,d\tau
\end{align*}
where $\hat{a}(\omega)$ and $\hat{a}(\tau)$ refer to the incident
wave, $\hat{b}(\omega)$ and $\hat{b}(t)$ refer to the reflected wave,
$\hat{S}(\omega)$ refers to the scattering matrix, and $\hat{S}(t)$
refers to the impulse-response matrix. Each pair is related via the
Fourier transforms. In a quality model, $\hat{S}(\omega)$ must be
reciprocal, passive, causal, have sampling density adequate to resolve
all features, and be defined over the entire spectrum of interest
\cite{vauinshteuin1976propagation, triverio2007stability}.

We apply the above method to a PTL with length $L$ and transmission
coefficient $S_{21}(\omega)=e^{-\gamma L}$, where $e^{-\gamma L}$ is
the propagation factor, $\gamma(\omega)=\alpha+i\beta$ is the
propagation constant, $\alpha$ is the attenuation constant, and
$\beta$ is the phase constant. The PTL input and output voltage in the
frequency domain relate as
\[
V_{\mbox{\footnotesize out}}(\omega)=V_{\mbox{\footnotesize in}}(\omega)\,e^{-\gamma L}
\] 
The Fourier transform of the input transient voltage
$V_{\mbox{\footnotesize in}}(t)$ is $V_{\mbox{\footnotesize
    in}}(\omega)$; the inverse Fourier transform of
$V_{\mbox{\footnotesize out}}(\omega)$ yields the output transient
$V_{\mbox{\footnotesize out}}(t)$.

\begin{figure}
 \centering
 \includegraphics[width=3.7in]{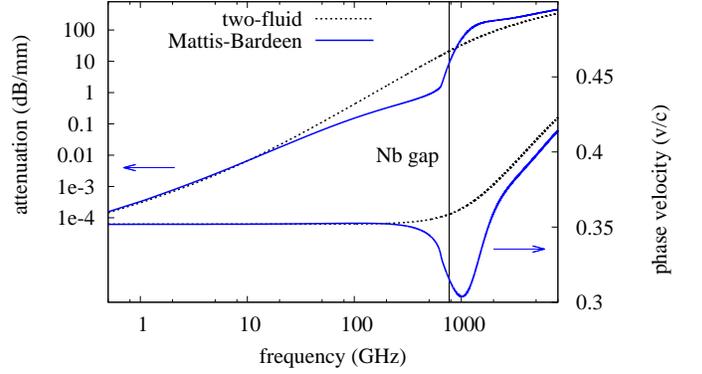}
 \caption{Comparison of the frequency-dependent attenuation and phase
   velocity between Mattis-Bardeen and the two-fluid model simulated
   for Nb wide microstrip line. Phase velocity is normalized to the
   speed of light, $c$. The models are aligned at 10\,GHz.
 \label{tf_mb}}
\end{figure}

To implement the time-domain model, the PTL propagation factor is used
to construct an S-parameter file to be used in the Spectre time-domain
circuit simulation. The full-length S-parameter data is calculated
during netlisting and is given as an input to a Verilog-A wrapper. The
model incorporates Spectre's ``nport'' device that provides the
impulse response. Spectre performs the inverse Fourier transform of
the scattering matrix
\[
\hat{S}_{\mbox{\footnotesize PTL}}(\omega)=\left[
  \begin{array}{cc}
    0           & 0 \\
    e^{-\gamma L} & 0
  \end{array}
  \right]
\]
to create the impulse response matrix $\hat{S}_{\mbox{\footnotesize
    PTL}}(t)$. The output transient is
\[
V_{\mbox{\footnotesize out}}(t)
=\int_{-\infty}^\infty V_{\mbox{\footnotesize in}}(\tau)\,
\hat{S}_{\mbox{\footnotesize PTL}}(t-\tau)\,d\tau
\]
which is the convolution of the impulse response with the input
signal, performed during the transient analysis.

\section{Frequency domain model and simulations}

To produce an analytical reference, we first calculate the complex
propagation constant $\gamma =\alpha +i\beta$ of a wide microstrip
formed by identical superconducting plates following the approach in
\cite{Talanov2000}:
\begin{equation}
  \beta =
  \omega\sqrt{\varepsilon_0\varepsilon_r\mu_0}
  \sqrt{1+\frac{2(X_{\mbox{\footnotesize eff}}
      -R_{\mbox{\footnotesize eff}}\tan{\delta})}{\omega\mu_0s}}
\label{EQ1beta}
\end{equation}

\begin{equation}
  \alpha =
  \frac{\beta}{2}
  \left[\frac{2 R_{\mbox{\footnotesize eff}}}
    {\omega\mu_0 s+2(X_{\mbox{\footnotesize eff}}-R_{\mbox{\footnotesize eff}}\tan{\delta})}
  +\tan{\delta}\right]
\label{EQ1alpha}
\end{equation}
where $\mu_0$ is the vacuum permeability, $s$ is the dielectric
thickness, $\tan{\delta}$ is the dielectric loss tangent,
$\varepsilon_0$ is the vacuum permittivity, $\varepsilon_r$ is the
interlayer dielectric constant, and it is assumed that $\alpha <<
\beta$. $R_{\mbox{\footnotesize eff}}$ and $X_{\mbox{\footnotesize
    eff}}$ are the real and imaginary parts of the effective surface
impedance of the superconducting plate of thickness $d$
\cite{kautz1978picosecond}\cite{klein1990effective}
\[
Z_{\mbox{\footnotesize eff}}
= R_{\mbox{\footnotesize eff}}+iX_{\mbox{\footnotesize eff}}
= \sqrt{i\mu_0\omega/\sigma}\,\coth\!\left(\sqrt{i\mu_0\omega\sigma}\,d\right)
\]

\begin{figure}
 \centering
 \includegraphics[width=3.0in]{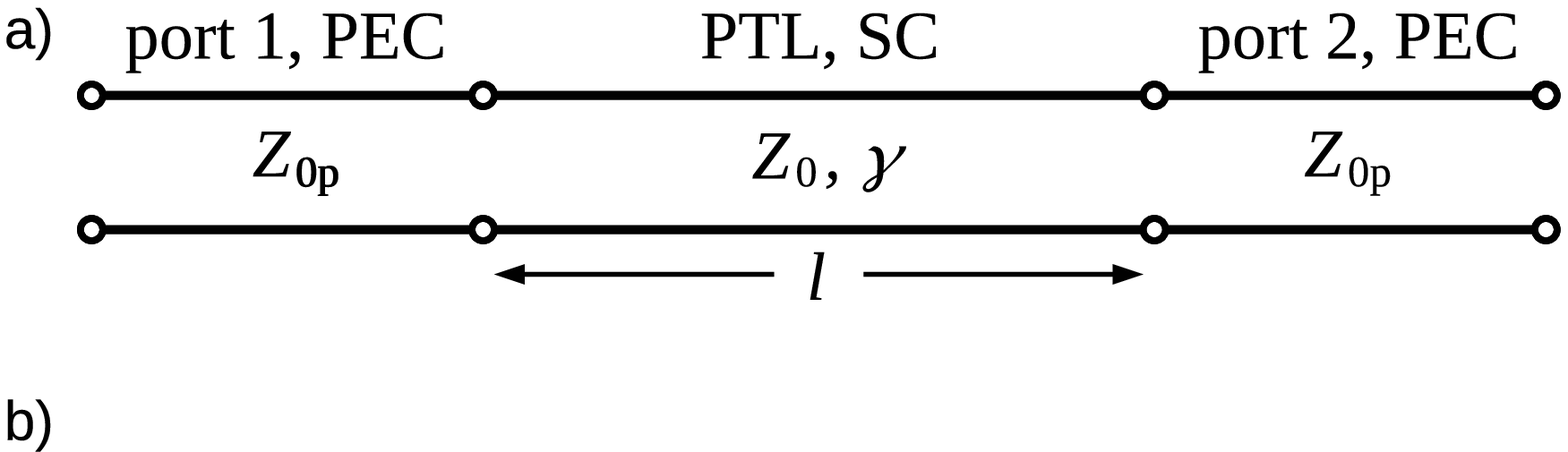}
 \includegraphics[width=2.8in]{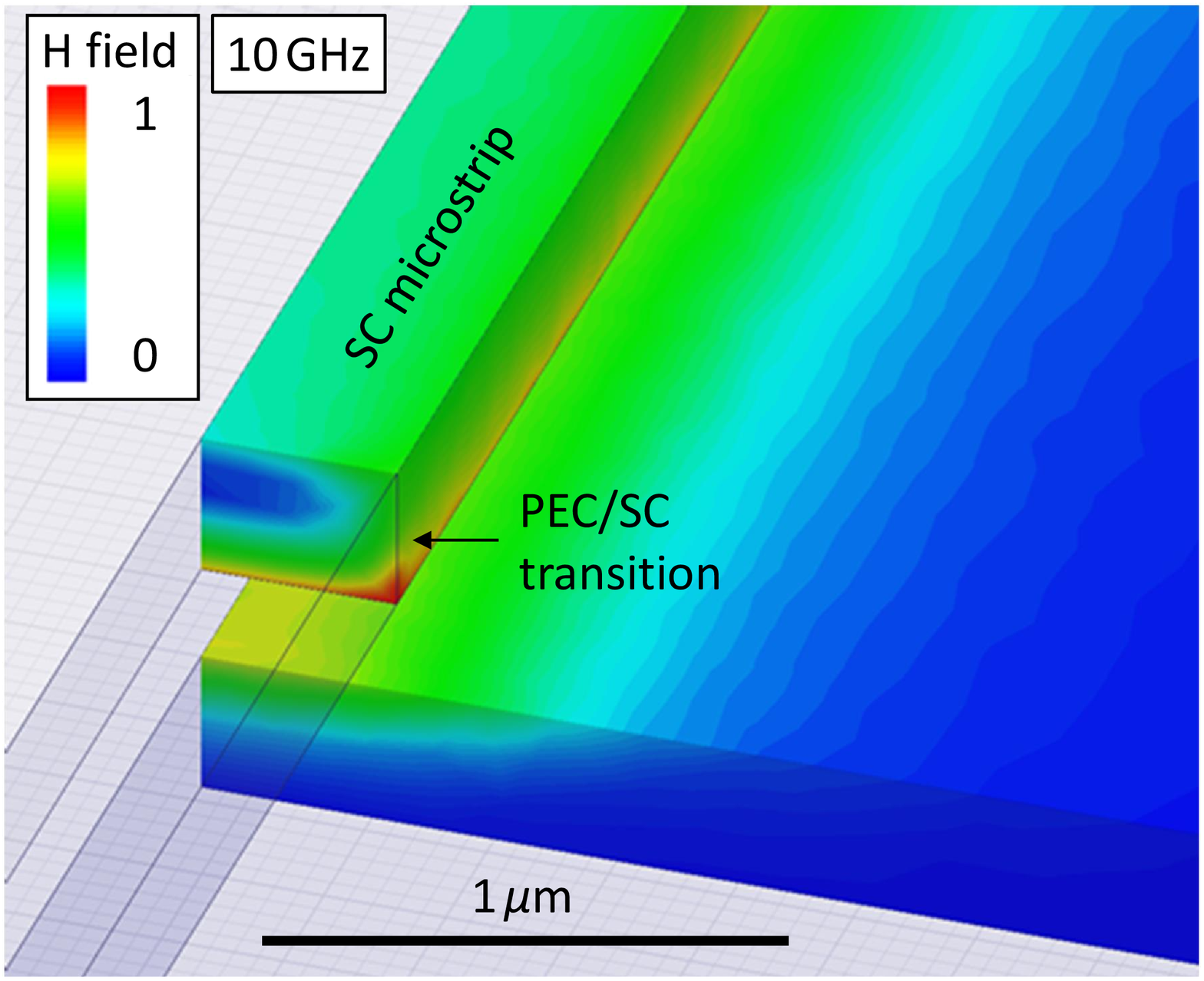}
 \caption{a) Schematic and b) HFSS model for a 1\,$\mu$m wide
   superconducting PTL. The simulation is in half-space, exploiting
   bilaterial symmetry. The length $l$ is 100-200\,$\mu$m. The
   magnetic field intensity in the superconductor PTL is shown at the
   port-PTL interface. The impedance mismatch between the PEC port
   with characteristic impedance $Z_{0p}$ and the superconductor PTL
   with characteristic impedance $Z_0$ causes a slight non-uniformity
   in the magnetic field near the port-PTL interface. In this
   simulation Nb thickness is 200\,nm and dielectric thickness is
   150\,nm. A dielectric material model with $\varepsilon_r$ of 4.2 and
   $\tan\delta$ of $10^{-3}$ are representative of the SiO$_2$
   interlayer dielectric used in superconducting processes
   \cite{Oates2017}.
 \label{schematic}}
\end{figure}

To simulate the PTLs in HFSS we use the Mattis-Bardeen microscopic
theory \cite{mattis1958theory} for the real and imaginary parts of Nb
complex conductivity $\sigma=\sigma_1-i\sigma_2$. Fig.~\ref{tf_mb}
shows the frequency-dependent attenuation and phase velocity of
Mattis-Bardeen model compared to the two-fluid model. The two-fluid
model overestimates attenuation and slightly underestimates dispersion
at frequencies approaching the gap. Overall, Mattis-Bardeen predicts
further propagation distance for the SFQ pulses. In the current
process the spectral width of the SFQ pulse is about 350\,GHz so we
expect the two-fluid quadratic dependence of propagation distance on
pulse width. In an advanced process with higher
critical-current-density junctions the difference between the two
models will be more significant.

In HFSS, the superconducting metal has been modeled as a material with
bulk conductivity $\sigma_1$ and negative relative permittivity
$-\sigma_2/\varepsilon_0\omega$. This model is based on the general
relationship between the complex permittivity and complex conductivity
of any material as $\varepsilon
=\varepsilon_0(\varepsilon_1-i\varepsilon_2) =-i\sigma/\omega$, where
$\varepsilon_1=-\sigma_2/\varepsilon_0\omega$ and $\varepsilon_2
=\sigma_1/\varepsilon_0\omega$ are the real and imaginary parts of the
relative permittivity. While we have modeled the superconductor as
having real bulk conductivity and negative relative permittivity, the
latest version of HFSS allows for a more direct approach by defining
materials with complex bulk conductivity.

\begin{figure}
 \centering
 \includegraphics[width=3.7in]{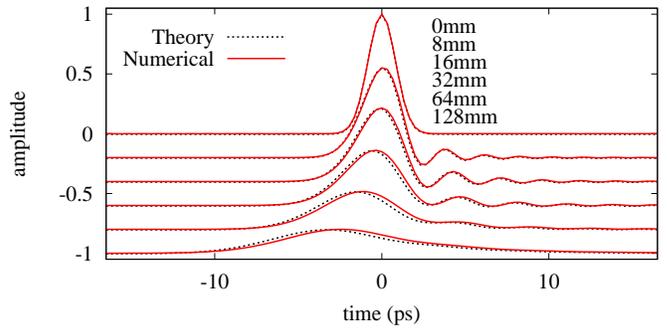}
 \caption{Pulse propagation on a superconducting passive transmission
   line. Subsequent traces are offset by -0.2\,mV for
   legibility. ``Theory'' is an analytical solution using Eq. (1-2),
   and ``Numerical'' is HFSS simulation. Geometrical and material
   parameters are similar except for PTL width. A 4\,$\mu$m width in
   HFSS simulations was chosen to produce a small fringe-field effect,
   giving good agreement with the analytical theory.
 \label{fig3}}
\end{figure}

Mattis-Bardeen theory offers no analytical expressions for the
superconductor complex conductivity, so we use HFSS datasets for
piece-wise representation of Nb material parameters. We employ the
theory to tabulate the Nb bulk conductivity and relative permittivity
vs.\ frequency for 1-1000\,GHz at 1000 frequency points, and import
the datasets into HFSS. The Mattis-Bardeen theory parameters which are
superconductor normal conductivity and energy gap, are chosen to give
the penetration depth
$\lambda=\mbox{Im}\left[\sqrt{i\mu_0\omega/\sigma}\right]/\mu_0\omega=90$\,nm
and to give the intrinsic surface resistance
$R_s=\mbox{Re}\left[\sqrt{i\mu_0\omega/\sigma}\right]=20$\,$\mu$$\Omega$
at 10\,GHz and 4.2\,K. These numbers are representative of Nb
interconnect with $T_c$ of 9\,K.

Fig.~\ref{schematic} shows a schematic and HFSS 3D model for a PTL
terminated with Perfect Conductor (PEC) ports of the same
cross-sectional geometry, as HFSS does not allow superconductor
ports. The perfect-conductor model has zero resistance and zero kinetic
inductance, so current flows only on the surface. After deembedding
the ports, the propagation factor for the PTL is computed as
\begin{equation}
  e^{-\gamma l}=\frac{1-S^2_{11}+S^2_{21}\pm
    \sqrt{\left(1-S^2_{11}+S^2_{21}\right)^2-4S^2_{21}}}{2S_{21}}
  \label{trustme}
\end{equation}
where $S_{ij}$ are the elements of the scattering matrix found by the
HFSS solver, and $l$ is the length of the simulated PTL. A deriviation
can be found in the Appendix. The requirement that the real and
imaginary parts of the propagation constant be positive defines the
sign of the square root. It is tractable for simulations to calculate
100-200 points for the propagation factor for frequencies of
1-1000\,GHz.

\begin{figure*}
 \centering
 \includegraphics[width=7.35in]{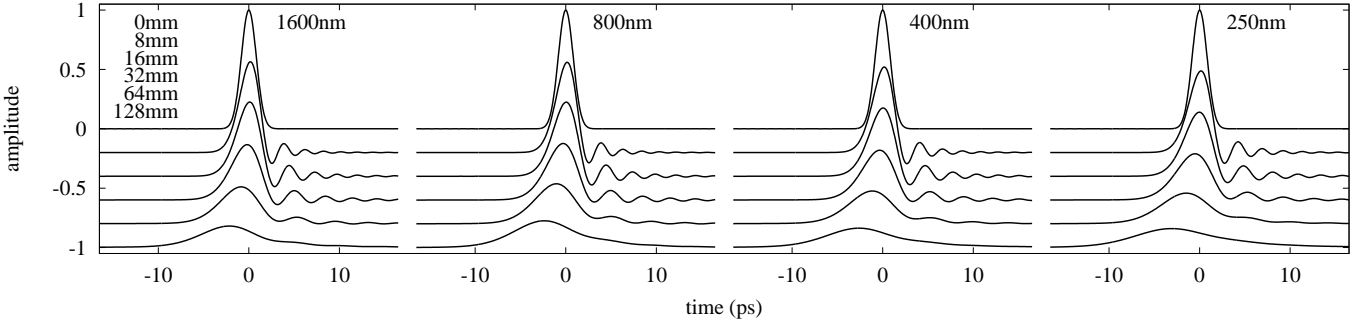}
 \caption{HFSS simulation of SFQ pulse propagation on a
   superconducting PTL for four microstrip widths, each at six
   lengths. Subsequent traces are again offset by -0.2\,mV for
   legibility. Dielectric thickness is 200\,nm, Nb microstrip and
   ground plane thickness is 300\,nm, the dielectric constant is 5.65,
   $\tan\delta$ is 0.0005, the Nb intrinsic surface resistance, $R_s$,
   is 20\,$\mu\Omega$ at 10\,GHz, and the penetration depth $\lambda$
   is 90\,nm.
 \label{fig4}}
\end{figure*}

The HFSS engine limits the aspect ratio of any feature to about
10,000:1, which limits the PTL model to a few mm in length. To extend
the model to arbitrary length, $L$, we mathematically construct
$e^{-\gamma L}$ from the simulated propagation factor $e^{-\gamma l}$
simulated for the PTL of length $l$. Implementation details of our
procedure include: 1) The function $\mbox{Arg}[e^{-\gamma l}]$ returns
the phase angle in the range $\pm \pi$.  The discontinuities are
removed to recover true continuous phase using e.g.\ MATLAB's
``unwrap'' function. 2) A virtual point at zero-frequency is added
with appropriate phase and amplitude corresponding to zero
attenuation. 3) Both magnitude and continuous phase are interpolated
between the simulated frequency points using piece-wise functions
$A(\omega)$ and $\theta(\omega)$:
\begin{align*}
  \lvert e^{-\gamma l} \rvert & \rightarrow A(\omega), &
  \mbox{unwrap}\left(\mbox{Arg}[e^{-\gamma l}]\right) \rightarrow \theta (\omega)
\end{align*}
where the arrows denote the interpolation. Finally, we compute the
full-length propagation factor
\[
  e^{-\gamma(\omega) L}=A(\omega)^{L/l}e^{i\theta(\omega) L/l}
\]

We first compare the simulation result for propagation of an SFQ pulse
to the analytical solution using Eqns.~\ref{EQ1beta} and
\ref{EQ1alpha} as shown in Fig.~\ref{fig3}. The SFQ pulse is
approximated by a Gaussian with $V_{\mbox{\footnotesize in}}(t)=V_0
e^{-t^2/2\tau^2}$, where the pulse magnitude $V_0$ of 1\,mV and FWHM
$\tau$ of 1.88\,ps produce $1\,\Phi_0$. The simulation and numerical
results agree, and both show significant dispersion at the longer
lengths.

Fig.~\ref{fig4} shows pulse propagation in microstrips of length
8-64\,mm as a function of width in the range 0.25-1.6\,$\mu$m. Pulse
amplitude decreases with length and is half the initial value after
propagating 64\,mm. The narrower, higher impedance lines have more
attentuation and dispersion due to increased kinetic inductance and
the dominant contribution of metalic loss. For a given impedance,
longer-range propagation can be achieved using a wider line and
thicker dielectric.

\section{Circuit Design and Measurement}

\begin{figure}
 \centering
 \includegraphics[width=3.5in]{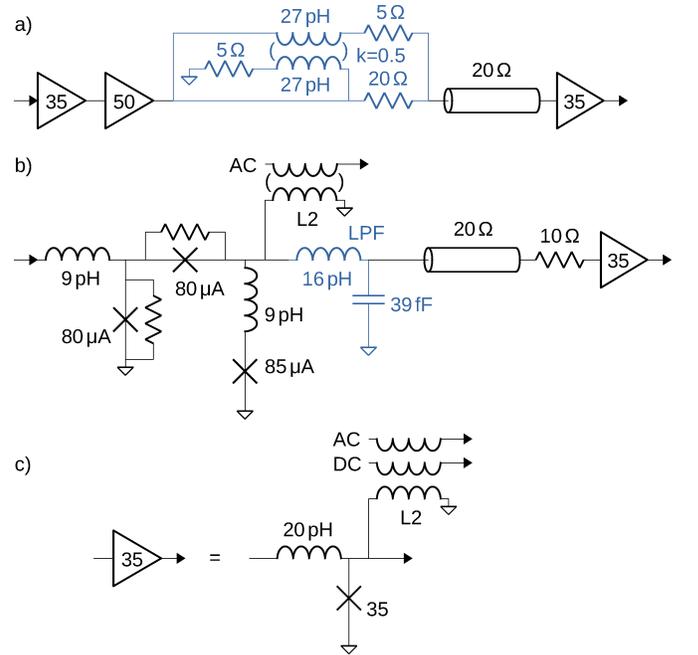}
 \caption{Interconnect schematics. a) The SFQ driver launchs a full
   SFQ pulse onto the transmission line through a network that both
   steps up the voltage and provides source termination. b) The DFQ
   driver launches a pulse that is about the same peak amplitude, but
   twice as wide, as the SFQ. A low-pass filter (LPF) shapes the
   pulse. The receiver absorbs half of the signal---one SFQ---while
   the other half is resistively terminated. c) The standard JTL for a
   35\,$\mu$A junction. The leading inductor has a reduced value of
   10\,pH in the receiver JTL.
 \label{schem}}
\end{figure}

\begin{figure}
 \centering
 \includegraphics[width=3.7in]{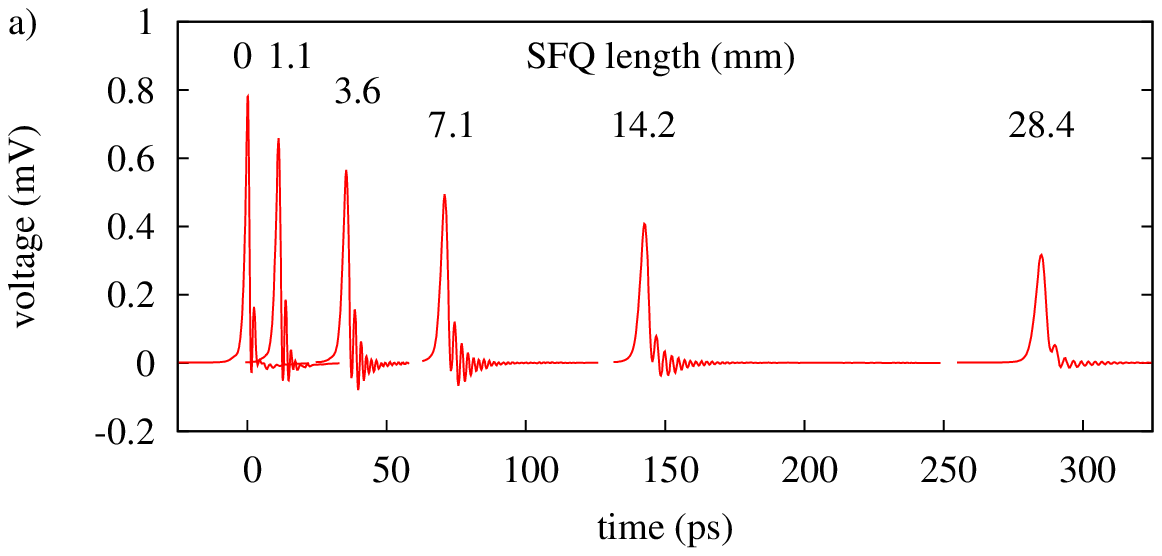}
 \includegraphics[width=3.7in]{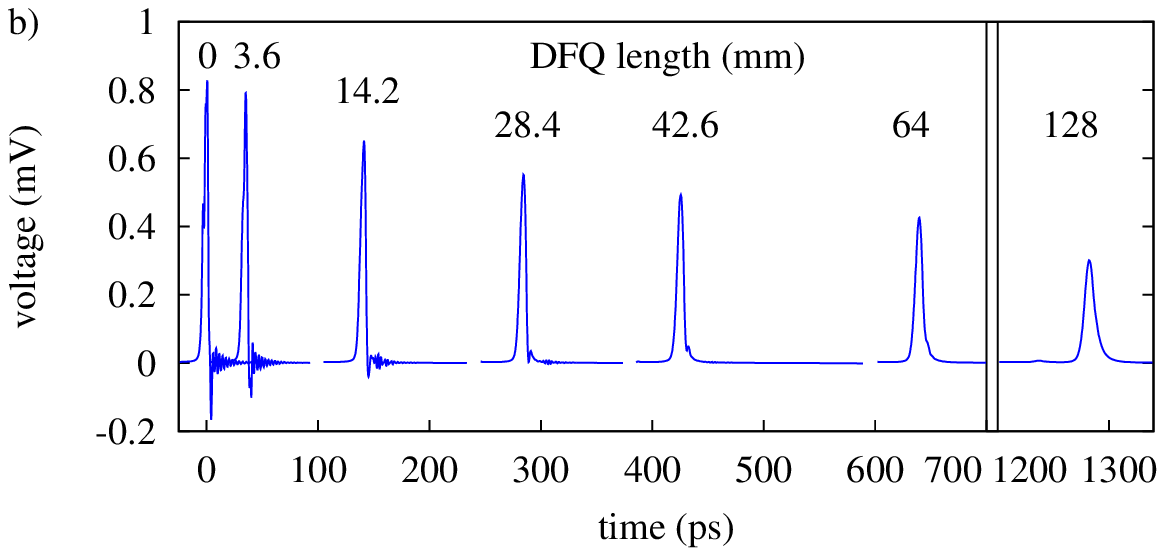}
 \includegraphics[width=3.7in]{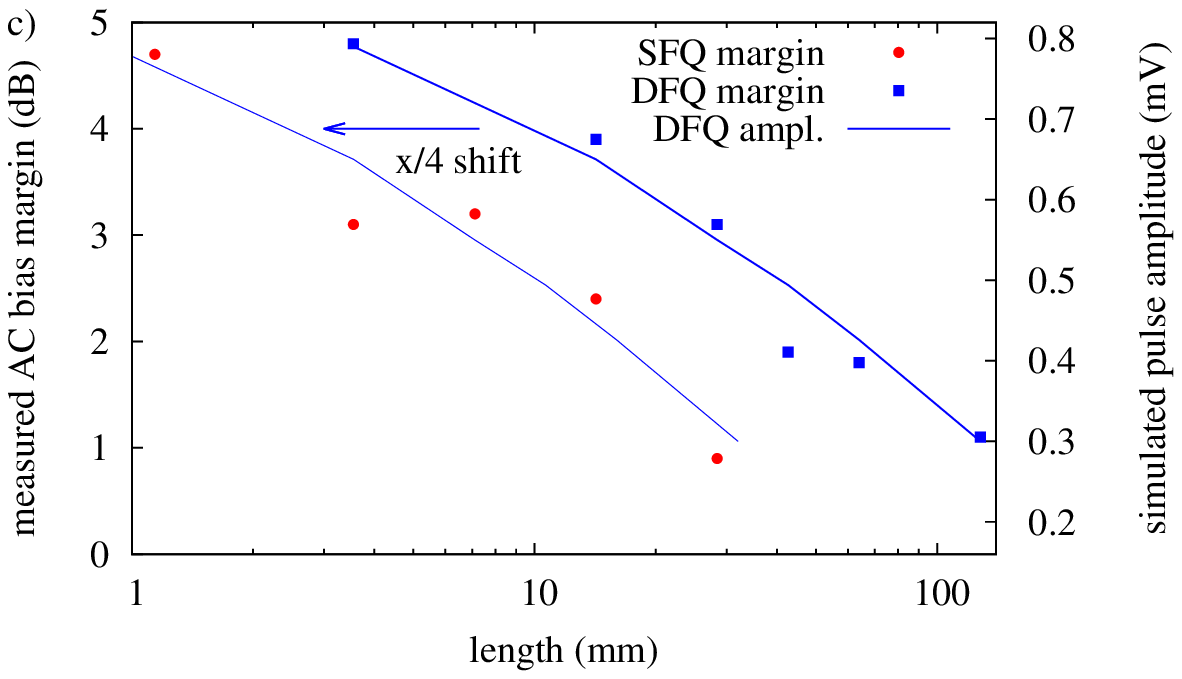}
 \caption{Time domain Spectre simulations of a) SFQ and b) DFQ pulse
   propagation for various transmission line lengths. The earliest
   waveform is taken at the driver. Subsequent waveforms are taken at
   the end of resistevely terminated PTLs of given lengths. c)
   Measured AC bias margin (points) for the various SFQ and DFQ
   transmission line lengths using a ``11110000111000110010'' chirp
   pattern at 3.5 GHz, with a signal-averaged output waveform. Peak
   pulse amplitude read from the DFQ simulations are overlaid with the
   measured margins, using a linear fit given by the scale and offset
   on the right. The same curve is shown shifted to the left by a
   factor of 1/4.
 \label{measure}}
\end{figure}

To verify our model we designed test circuits with different PTL
lengths and two different PTL drivers that generated SFQ and
double-flux-quantum (DFQ) pulses. Each PTL link consists of a driver,
a 20\,$\Omega$ PTL stripline, and a receiver. The driver is fed with a
differential input transformer \cite{herr2011ultra} and Josephson
Transmission Line (JTL) chain, and the receiver connects to a 2\,mV
differential amplifier at the output. Circuits are powered using a
resonant clock network at 3.5\,GHz that provisions the active area of
the chip \cite{talanov2017clock}.

PTL links were characterized with pulses having two different spectral
bandwidths, generated by two different PTL drivers. The DFQ driver
produces a pulse with twice the width of the SFQ, and similar peak
amplitude. Due to the quadratic dependence of attenuation constant
vs.\ frequency discussed above, the DFQ pulse is expected to propagate
four times the distance. Schematics of the drivers are shown in
Fig.~\ref{schem}.

The designs used effective terminations of 5-10\,$\Omega$ to damp
reflections. The SFQ driver used source termination. If the driver JTL
were connected directly to a series resistor, only a fraction of an
SFQ pulse would be launched on the PTL. The transformer between the
JTL and the PTL steps up the voltage and launches a full SFQ.  The
20\,$\Omega$ resistor bypasses the transformer to damp high
frequencies.

The DFQ driver produces two SFQ pulses, as described in
\cite{herr2002high}. In the current implementation, the output
terminal connects the central node of the driver, and the output
signal averages over the voltages produced among the junctions. The
two SFQ pulses partially overlap, and a simple LC low-pass filter on
the output produces a suitably unitary waveform.

The 20\,$\Omega$ PTL is matched to the characteristic impedance of
the 35\,$\mu$A receiver junction with critical current density
100\,$\mu$A/$\mu$m$^2$. The receiver junction absorbs an SFQ pulse when it
triggers. For the DFQ design, this corresponds to only half the
signal, and the other half is terminated in the 10\,$\Omega$ resistor
\cite{herr2003ballistic}.

Hitting the timing window of the AC-powered receiver is an important
design constraint. As the resonant clock network provides a global
timing reference, the length of the PTL in each circuit is an integer
number of $\lambda/4$ segments, where $\lambda$ is the wavelength of the
clock frequency. With appropriate phase selection at driver
and reciever, the arrival time at the receiver is centered in the
window for all PTL lengths.

The circuits were simulated in the time-domain with PTL models
generated for the specific geometry used, described below. Simulations
showed no degradation of margins due to reflections, at all
lengths. The design margin for AC power is 5\,dB ($\pm$28\%) for the
short PTLs where attenuation and dispersion are small. The circuits
included PTLs long enough to show margin degradation. The simulated
dispersion of the data pulses as a function of PTL length is shown in
Fig.~\ref{measure}a and b. These simulations also show four times
longer propagation distance for the DFQ pulses, based on a comparison
of peak amplitude at the longer lengths.

The circuits were fabricated in a process supplied by D-Wave
\cite{berkley2010scalable}, with six metal layers including the ground
plane, Josephson junctions with 100\,$\mu$A/$\mu$m$^2$ critical
current density, and with 0.25\,$\mu$m minimum feature size.  The PTLs
are implemented as stripline, with the signal layer between two ground
layers connected with side-wall vias. The 20\,$\Omega$ line was
optimized in HFSS using the design-rule metal and dielectric
thicknesses of 200\,nm and $\lambda_L$ of 100\,nm. The resulting
geometry is a 0.950\,$\mu$m wire width and 0.675\,$\mu$m spacing to
the via wall. PTL design is supported by the process design kit (PDK)
environment using the Multi-Part-Path (MPP) parameteric cell in
Cadence Virtuoso, and verification of connectivity and length using
Cadence Assura layout-versus-schematic (LVS).

The speed of light in the PTL was verified using S-parameter
measurements of a $\lambda/2$ resonator on the same chip and using the
same design as the PTLs of the functional circuits. The measured
velocity of 99.5\,$\mu$m/ps is within 3\% of the design
value, indicating sufficent fabrication process targeting.

Fig.~\ref{measure}c shows the measured AC bias margin as a function of
PTL length for both SFQ and DFQ pulses. AC bias provides a measure of
pulse amplitude at the receiver by changing receiver sensitivity. This
global AC bias also influences the amplitude of the pulse generated by
the driver, but this is a smaller effect due to SFQ pulse
quantization. The measured AC margin for DFQ pulses is overlayed with
simulated DFQ pulse height using data from Fig.~\ref{measure}b with a
fitted scale and offset. The same DFQ pulse height curve, shifted to
the left by a factor of 1/4, overlays the SFQ data. This confirms the
expected quadratic dependence of propagation distance on pulse width
and the correctness of the design and simulation methodology. Both
drivers show reduction of bias margin for longer PTLs. For practical
purposes the accepatble length of the PTL is defined by the AC bias
margin greater than 3\,dB that corresponds to the PTL lengths of 7\,mm
and 28\,mm for the two drivers.

\section{Conclusion}
Superconducting interconnects of arbitrary geometry and length can be
simulated in HFSS using the Mattis-Bardeen model. We have introduced a
method to simulate propagation of ps pulses over supercondutcing PTL
interconnects, based on HFSS simulations that provide an S-parameter
model for time-domain simulations in Spectre. Using this method we
have shown that the propagation distance is constrained by the high
frequency components of the pulse spectrum and by the PTL geometry.

We have shown in simulation and test that 20\,$\Omega$ Nb-based PTL
interconnect with on-chip dielectric thickness of 200\,nm can cover a
distance of 7\,mm using the SFQ driver and 28\,mm using DFQ. Range
scales with the square of the pulse width. The thicker dielectric
associated with multichip packaging, and increased pulse width
corresponding to 10 SFQ pulses per bit, would extend the reach to
meters.

\section*{Appendix}
The $ABCD$ matrix for a lossy PTL can be found in
\cite{paul2007analysis}
\[
ABCD=\left[
  \begin{array}{cc}
    \cosh(\gamma l)           & Z_0\sinh(\gamma l) \\
    \sinh(\gamma l)/Z_0       & \cosh(\gamma l)
  \end{array}
  \right]
\]
where $\gamma$ is the PTL propagation constant and $Z_0$ is the PTL
characteristic impedance. Converting this into a scattering matrix
\cite{pozar2011microwave4} yields
\begin{align}
  \left\{
    \begin{array}{l}
      S_{11}=S_{22}= \frac{(Z_0^2-Z_{0p}^2)\sinh(\gamma l)}
      {2Z_0Z_{0p}\cosh(\gamma l)+(Z_0^2-Z_{0p}^2)\sinh(\gamma l)} \\
      S_{21}=S_{12}= \frac{2Z_0Z_{0p}}
      {2Z_0Z_{0p}\cosh(\gamma l)+(Z_0^2-Z_{0p}^2)\sinh(\gamma l)}
    \end{array}
    \right.
    \label{A1}
\end{align}
where $Z_{0p}$ is the characteristic impedance of the PEC port.
Substituting $\cosh (\gamma l)=(e^{\gamma l} + e^{-\gamma l})/2$ and
$\sinh (\gamma l)=(e^{\gamma l} - e^{-\gamma l})/2$ into
Eqns.~\ref{A1} and solving for $e^{-\gamma l}$ and $Z_0$ leads to an
expression for PTL characteristic impedance in terms of the
S-parameters
\[
Z_0=\mp Z_{0p}\frac{\sqrt{\left(1-S_{11}^2+S_{21}^2\right)^2-4S_{21}^2}}{(1-S_{11})^2-S_{21}^2}
\]
and also leads to Eqn.~\ref{trustme}.

\begin{acknowledgments}

The authors acknowledge Yamil Morales and Tim Crowley for assisting
with circuit test, Paul Chang for assisting with data reduction and
analysis, and Joshua Strong for valuable discussions and help with
HFSS simulations.

\end{acknowledgments}

\bibliography{ptl_preprint}

\begin{thebibliography}{31}
\expandafter\ifx\csname natexlab\endcsname\relax\def\natexlab#1{#1}\fi
\expandafter\ifx\csname bibnamefont\endcsname\relax
  \def\bibnamefont#1{#1}\fi
\expandafter\ifx\csname bibfnamefont\endcsname\relax
  \def\bibfnamefont#1{#1}\fi
\expandafter\ifx\csname citenamefont\endcsname\relax
  \def\citenamefont#1{#1}\fi
\expandafter\ifx\csname url\endcsname\relax
  \def\url#1{\texttt{#1}}\fi
\expandafter\ifx\csname urlprefix\endcsname\relax\def\urlprefix{URL }\fi
\providecommand{\bibinfo}[2]{#2}
\providecommand{\eprint}[2][]{\url{#2}}

\bibitem[{\citenamefont{Herr et~al.}(2002)\citenamefont{Herr, Smith, and
  Wire}}]{herr2002high}
\bibinfo{author}{\bibfnamefont{Q.~P.} \bibnamefont{Herr}},
  \bibinfo{author}{\bibfnamefont{A.~D.} \bibnamefont{Smith}}, \bibnamefont{and}
  \bibinfo{author}{\bibfnamefont{M.~S.} \bibnamefont{Wire}},
  \bibinfo{journal}{Applied Physics Letters} \textbf{\bibinfo{volume}{80}},
  \bibinfo{pages}{3210} (\bibinfo{year}{2002}).

\bibitem[{\citenamefont{Hashimoto et~al.}(2005)\citenamefont{Hashimoto, Yorozu,
  Satoh, and Miyazaki}}]{hashimoto2005demonstration}
\bibinfo{author}{\bibfnamefont{Y.}~\bibnamefont{Hashimoto}},
  \bibinfo{author}{\bibfnamefont{S.}~\bibnamefont{Yorozu}},
  \bibinfo{author}{\bibfnamefont{T.}~\bibnamefont{Satoh}}, \bibnamefont{and}
  \bibinfo{author}{\bibfnamefont{T.}~\bibnamefont{Miyazaki}},
  \bibinfo{journal}{Applied Physics Letters} \textbf{\bibinfo{volume}{87}},
  \bibinfo{pages}{022502} (\bibinfo{year}{2005}).

\bibitem[{\citenamefont{Filippov et~al.}(2017)\citenamefont{Filippov, Amparo,
  Kamkar, Walter, Kirichenko, Mukhanov, and Vernik}}]{filippov2017experimental}
\bibinfo{author}{\bibfnamefont{T.}~\bibnamefont{Filippov}},
  \bibinfo{author}{\bibfnamefont{D.}~\bibnamefont{Amparo}},
  \bibinfo{author}{\bibfnamefont{M.}~\bibnamefont{Kamkar}},
  \bibinfo{author}{\bibfnamefont{J.}~\bibnamefont{Walter}},
  \bibinfo{author}{\bibfnamefont{A.}~\bibnamefont{Kirichenko}},
  \bibinfo{author}{\bibfnamefont{O.}~\bibnamefont{Mukhanov}}, \bibnamefont{and}
  \bibinfo{author}{\bibfnamefont{I.}~\bibnamefont{Vernik}}, in
  \emph{\bibinfo{booktitle}{2017 16th International Superconductive Electronics
  Conference (ISEC)}} (\bibinfo{organization}{IEEE}, \bibinfo{year}{2017}), pp.
  \bibinfo{pages}{1--4}.

\bibitem[{\citenamefont{Farjadrad et~al.}(2019)\citenamefont{Farjadrad,
  Kuemerle, and Vinnakota}}]{farjadrad2019bunch}
\bibinfo{author}{\bibfnamefont{R.}~\bibnamefont{Farjadrad}},
  \bibinfo{author}{\bibfnamefont{M.}~\bibnamefont{Kuemerle}}, \bibnamefont{and}
  \bibinfo{author}{\bibfnamefont{B.}~\bibnamefont{Vinnakota}},
  \bibinfo{journal}{IEEE Micro} \textbf{\bibinfo{volume}{40}},
  \bibinfo{pages}{15} (\bibinfo{year}{2019}).

\bibitem[{\citenamefont{Miller}(2017)}]{miller2017attojoule}
\bibinfo{author}{\bibfnamefont{D.~A.} \bibnamefont{Miller}},
  \bibinfo{journal}{Journal of Lightwave Technology}
  \textbf{\bibinfo{volume}{35}}, \bibinfo{pages}{346} (\bibinfo{year}{2017}).

\bibitem[{\citenamefont{Egan and et. al.}(2021)}]{Egan2020}
\bibinfo{author}{\bibfnamefont{J.}~\bibnamefont{Egan}} \bibnamefont{and}
  \bibinfo{author}{\bibnamefont{et. al.}}, \bibinfo{journal}{to be published}
  (\bibinfo{year}{2021}).

\bibitem[{\citenamefont{Herr et~al.}(2011)\citenamefont{Herr, Herr, Oberg, and
  Ioannidis}}]{herr2011ultra}
\bibinfo{author}{\bibfnamefont{Q.~P.} \bibnamefont{Herr}},
  \bibinfo{author}{\bibfnamefont{A.~Y.} \bibnamefont{Herr}},
  \bibinfo{author}{\bibfnamefont{O.~T.} \bibnamefont{Oberg}}, \bibnamefont{and}
  \bibinfo{author}{\bibfnamefont{A.~G.} \bibnamefont{Ioannidis}},
  \bibinfo{journal}{Journal of Applied Physics} \textbf{\bibinfo{volume}{109}},
  \bibinfo{pages}{103903} (\bibinfo{year}{2011}).

\bibitem[{\citenamefont{{Volkmann} et~al.}(2015)\citenamefont{{Volkmann},
  {Vernik}, and {Mukhanov}}}]{Volkmann2015}
\bibinfo{author}{\bibfnamefont{M.~H.} \bibnamefont{{Volkmann}}},
  \bibinfo{author}{\bibfnamefont{I.~V.} \bibnamefont{{Vernik}}},
  \bibnamefont{and} \bibinfo{author}{\bibfnamefont{O.~A.}
  \bibnamefont{{Mukhanov}}}, \bibinfo{journal}{IEEE Transactions on Applied
  Superconductivity} \textbf{\bibinfo{volume}{25}}, \bibinfo{pages}{1}
  (\bibinfo{year}{2015}).

\bibitem[{\citenamefont{China et~al.}(2016)\citenamefont{China, Tsuji, Narama,
  Takeuchi, Ortlepp, Yamanashi, and Yoshikawa}}]{china2016demonstration}
\bibinfo{author}{\bibfnamefont{F.}~\bibnamefont{China}},
  \bibinfo{author}{\bibfnamefont{N.}~\bibnamefont{Tsuji}},
  \bibinfo{author}{\bibfnamefont{T.}~\bibnamefont{Narama}},
  \bibinfo{author}{\bibfnamefont{N.}~\bibnamefont{Takeuchi}},
  \bibinfo{author}{\bibfnamefont{T.}~\bibnamefont{Ortlepp}},
  \bibinfo{author}{\bibfnamefont{Y.}~\bibnamefont{Yamanashi}},
  \bibnamefont{and}
  \bibinfo{author}{\bibfnamefont{N.}~\bibnamefont{Yoshikawa}},
  \bibinfo{journal}{IEEE Transactions on Applied Superconductivity}
  \textbf{\bibinfo{volume}{27}}, \bibinfo{pages}{1} (\bibinfo{year}{2016}).

\bibitem[{\citenamefont{Rafique et~al.}(2005)\citenamefont{Rafique, Kataeva,
  Engseth, Tarasov, and Kidiyarova-Shevchenko}}]{rafique2005optimization}
\bibinfo{author}{\bibfnamefont{M.}~\bibnamefont{Rafique}},
  \bibinfo{author}{\bibfnamefont{I.}~\bibnamefont{Kataeva}},
  \bibinfo{author}{\bibfnamefont{H.}~\bibnamefont{Engseth}},
  \bibinfo{author}{\bibfnamefont{M.}~\bibnamefont{Tarasov}}, \bibnamefont{and}
  \bibinfo{author}{\bibfnamefont{A.}~\bibnamefont{Kidiyarova-Shevchenko}},
  \bibinfo{journal}{Superconductor Science and Technology}
  \textbf{\bibinfo{volume}{18}}, \bibinfo{pages}{1065} (\bibinfo{year}{2005}).

\bibitem[{\citenamefont{Takeuchi et~al.}(2009)\citenamefont{Takeuchi,
  Yamanashi, Saito, and Yoshikawa}}]{TAKEUCHI2009}
\bibinfo{author}{\bibfnamefont{N.}~\bibnamefont{Takeuchi}},
  \bibinfo{author}{\bibfnamefont{Y.}~\bibnamefont{Yamanashi}},
  \bibinfo{author}{\bibfnamefont{Y.}~\bibnamefont{Saito}}, \bibnamefont{and}
  \bibinfo{author}{\bibfnamefont{N.}~\bibnamefont{Yoshikawa}},
  \bibinfo{journal}{Physica C: Superconductivity}
  \textbf{\bibinfo{volume}{469}}, \bibinfo{pages}{1662} (\bibinfo{year}{2009}).

\bibitem[{\citenamefont{{Polonsky} et~al.}(1993)\citenamefont{{Polonsky},
  {Semenov}, and {Schneider}}}]{Semenov93}
\bibinfo{author}{\bibfnamefont{S.~V.} \bibnamefont{{Polonsky}}},
  \bibinfo{author}{\bibfnamefont{V.~K.} \bibnamefont{{Semenov}}},
  \bibnamefont{and} \bibinfo{author}{\bibfnamefont{D.~F.}
  \bibnamefont{{Schneider}}}, \bibinfo{journal}{IEEE Transactions on Applied
  Superconductivity} \textbf{\bibinfo{volume}{3}}, \bibinfo{pages}{2598}
  (\bibinfo{year}{1993}).

\bibitem[{\citenamefont{{Takagi} et~al.}(2009)\citenamefont{{Takagi}, {Tanaka},
  {Iwasaki}, {Kasagi}, {Kataeva}, {Nagasawa}, {Satoh}, {Akaike}, and
  {Fujimaki}}}]{TanakaPTL2009}
\bibinfo{author}{\bibfnamefont{K.}~\bibnamefont{{Takagi}}},
  \bibinfo{author}{\bibfnamefont{M.}~\bibnamefont{{Tanaka}}},
  \bibinfo{author}{\bibfnamefont{S.}~\bibnamefont{{Iwasaki}}},
  \bibinfo{author}{\bibfnamefont{R.}~\bibnamefont{{Kasagi}}},
  \bibinfo{author}{\bibfnamefont{I.}~\bibnamefont{{Kataeva}}},
  \bibinfo{author}{\bibfnamefont{S.}~\bibnamefont{{Nagasawa}}},
  \bibinfo{author}{\bibfnamefont{T.}~\bibnamefont{{Satoh}}},
  \bibinfo{author}{\bibfnamefont{H.}~\bibnamefont{{Akaike}}}, \bibnamefont{and}
  \bibinfo{author}{\bibfnamefont{A.}~\bibnamefont{{Fujimaki}}},
  \bibinfo{journal}{IEEE Transactions on Applied Superconductivity}
  \textbf{\bibinfo{volume}{19}}, \bibinfo{pages}{617} (\bibinfo{year}{2009}).

\bibitem[{\citenamefont{{Das} et~al.}(2017)\citenamefont{{Das}, {Bolkhovsky},
  {Tolpygo}, {Gouker}, {Johnson}, {Dauler}, and {Gouker}}}]{LincolnMCM}
\bibinfo{author}{\bibfnamefont{R.~N.} \bibnamefont{{Das}}},
  \bibinfo{author}{\bibfnamefont{V.}~\bibnamefont{{Bolkhovsky}}},
  \bibinfo{author}{\bibfnamefont{S.~K.} \bibnamefont{{Tolpygo}}},
  \bibinfo{author}{\bibfnamefont{P.}~\bibnamefont{{Gouker}}},
  \bibinfo{author}{\bibfnamefont{L.~M.} \bibnamefont{{Johnson}}},
  \bibinfo{author}{\bibfnamefont{E.~A.} \bibnamefont{{Dauler}}},
  \bibnamefont{and} \bibinfo{author}{\bibfnamefont{M.~A.}
  \bibnamefont{{Gouker}}}, in \emph{\bibinfo{booktitle}{2017 IEEE 67th
  Electronic Components and Technology Conference (ECTC)}}
  (\bibinfo{year}{2017}), pp. \bibinfo{pages}{675--683}.

\bibitem[{\citenamefont{Shukla et~al.}(2019)\citenamefont{Shukla, Chonigman,
  Sahu, Kirichenko, Inamdar, and Gupta}}]{shukla2019investigation}
\bibinfo{author}{\bibfnamefont{A.}~\bibnamefont{Shukla}},
  \bibinfo{author}{\bibfnamefont{B.}~\bibnamefont{Chonigman}},
  \bibinfo{author}{\bibfnamefont{A.}~\bibnamefont{Sahu}},
  \bibinfo{author}{\bibfnamefont{D.}~\bibnamefont{Kirichenko}},
  \bibinfo{author}{\bibfnamefont{A.}~\bibnamefont{Inamdar}}, \bibnamefont{and}
  \bibinfo{author}{\bibfnamefont{D.}~\bibnamefont{Gupta}},
  \bibinfo{journal}{IEEE Transactions on Applied Superconductivity}
  \textbf{\bibinfo{volume}{29}}, \bibinfo{pages}{1} (\bibinfo{year}{2019}).

\bibitem[{\citenamefont{Mattis and Bardeen}(1958)}]{mattis1958theory}
\bibinfo{author}{\bibfnamefont{D.}~\bibnamefont{Mattis}} \bibnamefont{and}
  \bibinfo{author}{\bibfnamefont{J.}~\bibnamefont{Bardeen}},
  \bibinfo{journal}{Physical Review} \textbf{\bibinfo{volume}{111}},
  \bibinfo{pages}{412} (\bibinfo{year}{1958}).

\bibitem[{\citenamefont{Kautz}(1978)}]{kautz1978picosecond}
\bibinfo{author}{\bibfnamefont{R.~L.} \bibnamefont{Kautz}},
  \bibinfo{journal}{Journal of Applied Physics} \textbf{\bibinfo{volume}{49}},
  \bibinfo{pages}{308} (\bibinfo{year}{1978}).

\bibitem[{\citenamefont{Peterson and McDonald}(1977)}]{peterson1977picosecond}
\bibinfo{author}{\bibfnamefont{R.}~\bibnamefont{Peterson}} \bibnamefont{and}
  \bibinfo{author}{\bibfnamefont{D.}~\bibnamefont{McDonald}},
  \bibinfo{journal}{IEEE Transactions on Magnetics}
  \textbf{\bibinfo{volume}{13}}, \bibinfo{pages}{887} (\bibinfo{year}{1977}).

\bibitem[{\citenamefont{Chi et~al.}(1987)\citenamefont{Chi, Gallagher, Duling,
  Grischkowsky, Halas, Ketchen, and Kleinsasser}}]{chi1987subpicosecond}
\bibinfo{author}{\bibfnamefont{C.}~\bibnamefont{Chi}},
  \bibinfo{author}{\bibfnamefont{W.}~\bibnamefont{Gallagher}},
  \bibinfo{author}{\bibfnamefont{I.}~\bibnamefont{Duling}},
  \bibinfo{author}{\bibfnamefont{D.}~\bibnamefont{Grischkowsky}},
  \bibinfo{author}{\bibfnamefont{N.}~\bibnamefont{Halas}},
  \bibinfo{author}{\bibfnamefont{M.}~\bibnamefont{Ketchen}}, \bibnamefont{and}
  \bibinfo{author}{\bibfnamefont{A.}~\bibnamefont{Kleinsasser}},
  \bibinfo{journal}{IEEE Transactions on Magnetics}
  \textbf{\bibinfo{volume}{23}}, \bibinfo{pages}{1666} (\bibinfo{year}{1987}).

\bibitem[{HFS()}]{HFSS}
\emph{\bibinfo{title}{{3D} electromagnetic field simulator for {RF} and
  wireless design}},
  \bibinfo{note}{https://www.ansys.com/products/electronics/ansys-hfss}.

\bibitem[{\citenamefont{Shlepnev}(2010)}]{shlepnev2010quality}
\bibinfo{author}{\bibfnamefont{Y.}~\bibnamefont{Shlepnev}}, in
  \emph{\bibinfo{booktitle}{DesignCon IBIS Summit}} (\bibinfo{year}{2010}).

\bibitem[{\citenamefont{Vainshtein}(1976)}]{vauinshteuin1976propagation}
\bibinfo{author}{\bibfnamefont{L.~A.} \bibnamefont{Vainshtein}},
  \bibinfo{journal}{Soviet Physics Uspekhi} \textbf{\bibinfo{volume}{19}},
  \bibinfo{pages}{189} (\bibinfo{year}{1976}).

\bibitem[{\citenamefont{Triverio et~al.}(2007)\citenamefont{Triverio,
  Grivet-Talocia, Nakhla, Canavero, and Achar}}]{triverio2007stability}
\bibinfo{author}{\bibfnamefont{P.}~\bibnamefont{Triverio}},
  \bibinfo{author}{\bibfnamefont{S.}~\bibnamefont{Grivet-Talocia}},
  \bibinfo{author}{\bibfnamefont{M.~S.} \bibnamefont{Nakhla}},
  \bibinfo{author}{\bibfnamefont{F.~G.} \bibnamefont{Canavero}},
  \bibnamefont{and} \bibinfo{author}{\bibfnamefont{R.}~\bibnamefont{Achar}},
  \bibinfo{journal}{IEEE Transactions on Advanced Packaging}
  \textbf{\bibinfo{volume}{30}}, \bibinfo{pages}{795} (\bibinfo{year}{2007}).

\bibitem[{\citenamefont{{Talanov} et~al.}(2000)\citenamefont{{Talanov},
  {Mercaldo}, and {Anlage}}}]{Talanov2000}
\bibinfo{author}{\bibfnamefont{V.~V.} \bibnamefont{{Talanov}}},
  \bibinfo{author}{\bibfnamefont{L.~V.} \bibnamefont{{Mercaldo}}},
  \bibnamefont{and} \bibinfo{author}{\bibfnamefont{S.~M.}
  \bibnamefont{{Anlage}}}, \bibinfo{journal}{Review of Scientific Instruments}
  \textbf{\bibinfo{volume}{71}}, \bibinfo{pages}{2136} (\bibinfo{year}{2000}).

\bibitem[{\citenamefont{Klein et~al.}(1990)\citenamefont{Klein, Chaloupka,
  M{\"u}ller, Orbach, Piel, Roas, Schultz, Klein, and
  Peiniger}}]{klein1990effective}
\bibinfo{author}{\bibfnamefont{N.}~\bibnamefont{Klein}},
  \bibinfo{author}{\bibfnamefont{H.}~\bibnamefont{Chaloupka}},
  \bibinfo{author}{\bibfnamefont{G.}~\bibnamefont{M{\"u}ller}},
  \bibinfo{author}{\bibfnamefont{S.}~\bibnamefont{Orbach}},
  \bibinfo{author}{\bibfnamefont{H.}~\bibnamefont{Piel}},
  \bibinfo{author}{\bibfnamefont{B.}~\bibnamefont{Roas}},
  \bibinfo{author}{\bibfnamefont{L.}~\bibnamefont{Schultz}},
  \bibinfo{author}{\bibfnamefont{U.}~\bibnamefont{Klein}}, \bibnamefont{and}
  \bibinfo{author}{\bibfnamefont{M.}~\bibnamefont{Peiniger}},
  \bibinfo{journal}{Journal of Applied Physics} \textbf{\bibinfo{volume}{67}},
  \bibinfo{pages}{6940} (\bibinfo{year}{1990}).

\bibitem[{\citenamefont{{Oates} et~al.}(2017)\citenamefont{{Oates}, {Tolpygo},
  and {Bolkhovsky}}}]{Oates2017}
\bibinfo{author}{\bibfnamefont{D.~E.} \bibnamefont{{Oates}}},
  \bibinfo{author}{\bibfnamefont{S.~K.} \bibnamefont{{Tolpygo}}},
  \bibnamefont{and}
  \bibinfo{author}{\bibfnamefont{V.}~\bibnamefont{{Bolkhovsky}}},
  \bibinfo{journal}{IEEE Transactions on Applied Superconductivity}
  \textbf{\bibinfo{volume}{27}}, \bibinfo{pages}{1} (\bibinfo{year}{2017}).

\bibitem[{\citenamefont{Talanov and Strong}(2017)}]{talanov2017clock}
\bibinfo{author}{\bibfnamefont{V.~V.} \bibnamefont{Talanov}} \bibnamefont{and}
  \bibinfo{author}{\bibfnamefont{J.~A.} \bibnamefont{Strong}},
  \emph{\bibinfo{title}{Clock distribution network for a superconducting
  integrated circuit}} (\bibinfo{year}{2017}), \bibinfo{note}{{U}S Patent
  9,722,589}.

\bibitem[{\citenamefont{Herr et~al.}(2003)\citenamefont{Herr, Wire, and
  Smith}}]{herr2003ballistic}
\bibinfo{author}{\bibfnamefont{Q.~P.} \bibnamefont{Herr}},
  \bibinfo{author}{\bibfnamefont{M.~S.} \bibnamefont{Wire}}, \bibnamefont{and}
  \bibinfo{author}{\bibfnamefont{A.~D.} \bibnamefont{Smith}},
  \bibinfo{journal}{IEEE transactions on Applied Superconductivity}
  \textbf{\bibinfo{volume}{13}}, \bibinfo{pages}{463} (\bibinfo{year}{2003}).

\bibitem[{\citenamefont{Berkley et~al.}(2010)\citenamefont{Berkley, Johnson,
  Bunyk, Harris, Johansson, Lanting, Ladizinsky, Tolkacheva, Amin, and
  Rose}}]{berkley2010scalable}
\bibinfo{author}{\bibfnamefont{A.}~\bibnamefont{Berkley}},
  \bibinfo{author}{\bibfnamefont{M.}~\bibnamefont{Johnson}},
  \bibinfo{author}{\bibfnamefont{P.}~\bibnamefont{Bunyk}},
  \bibinfo{author}{\bibfnamefont{R.}~\bibnamefont{Harris}},
  \bibinfo{author}{\bibfnamefont{J.}~\bibnamefont{Johansson}},
  \bibinfo{author}{\bibfnamefont{T.}~\bibnamefont{Lanting}},
  \bibinfo{author}{\bibfnamefont{E.}~\bibnamefont{Ladizinsky}},
  \bibinfo{author}{\bibfnamefont{E.}~\bibnamefont{Tolkacheva}},
  \bibinfo{author}{\bibfnamefont{M.}~\bibnamefont{Amin}}, \bibnamefont{and}
  \bibinfo{author}{\bibfnamefont{G.}~\bibnamefont{Rose}},
  \bibinfo{journal}{Superconductor Science and Technology}
  \textbf{\bibinfo{volume}{23}}, \bibinfo{pages}{105014}
  (\bibinfo{year}{2010}).

\bibitem[{\citenamefont{Paul}(2007)}]{paul2007analysis}
\bibinfo{author}{\bibfnamefont{C.~R.} \bibnamefont{Paul}},
  \emph{\bibinfo{title}{Analysis of multiconductor transmission lines}}
  (\bibinfo{publisher}{John Wiley \& Sons}, \bibinfo{year}{2007}),
  chap.~\bibinfo{chapter}{6}.

\bibitem[{\citenamefont{Pozar}(2011)}]{pozar2011microwave4}
\bibinfo{author}{\bibfnamefont{D.~M.} \bibnamefont{Pozar}},
  \emph{\bibinfo{title}{Microwave engineering 4th edition}}
  (\bibinfo{publisher}{John Wiley \& Sons}, \bibinfo{year}{2011}),
  chap.~\bibinfo{chapter}{4}.

\end{thebibliography}

\end{document}